\documentstyle[preprint,aps]{revtex}
\begin{document}
\draft
\title{Why Nature has made 
a choice of  one time and three space coordinates?}
\author{ N. Manko\v c Bor\v stnik}
\address{ Department of Physics, University of
Ljubljana, Jadranska 19, 1111,
and Primorska Institute for Natural Sciences and Technology, 
C. Mare\v zganskega upora 2, Koper 6000, Slovenia}
\author{ H. B. Nielsen}
\address{Department of Physics, Niels Bohr Institute,
Blegdamsvej 17,\\
Copenhagen, DK-2100 }

\date{\today}

\maketitle

\begin{abstract} 
We  propose a possible answer to one of the most exciting open questions in physics and 
cosmology, that is  the question
why we seem to  experience four-dimensional space-time with three ordinary 
and one time dimensions.  

Making assumptions (such as particles being in first approximation massless) about 
the equations of motion, 
we argue for restrictions on the number of space 
and time dimensions.
(Actually the Standard model 
predicts and  experiments confirm that elementary particles are massless until interactions 
switch on masses.) 

Accepting our explanation of the space-time signature and the number of dimensions would be a point supporting 
(further) the importance of the ''internal space''.
\end{abstract}

\section{Introduction}\label{introduction}

There are many experiences, prejudices  so deeply embeded in our language 
and way of thinking, and so strongly connected with the idea that we have just one time dimension,  
that they could be used as arguments for such an idea, like the argument that the ordering ''before'' and ''after''
makes sense presupposing that there is just one time dimension.
However, all these concepts and prejudices and experiences do not 
constitute a genuine explanation of why we were placed into just such a world. 
The main point of the present work is to discuss a more microphysical 
explanation for features of the numbers of space and time dimensions\cite{hawking,hawking1,tegmark,%
holgernorma2000,%
ord,ord1,ord2,holger,holger1,wheeler,chew,woo,froggattnielsenbook,greensite93,%
greensite94,weinberg,penrose}
by associating these numbers
with properties of the equations - ''equations of motion'' - obeyed by the fields of the 
elementary particles, especially involving what we can call the ''internal space'', by which we mean 
the space of spins and charges and so that we can think of all the different particles, which are
fermions - like
quarks and leptons (and equivalently for bosons), as being different internal states of the same particle.

Theories of strings and membranes\cite{strings}  and Kaluza-Klein-like theories\cite{kaluzaklein} 
 (as well as the approach by one of us\cite{norma92,norma93,norma01},  which 
unifies spins and charges
in  the space of anticommuting coordinates and 
predicts the connection between the space-time dimension and the internal degrees of freedom) predict 
initially more-than-four-dimensional space-time. 
If this is true, how and when  did our Universe  in its evolution  choose the Minkowski metric,
and  when and in which way did it  ''decide'' to (''mostly'') manifest in four-dimensional 
space-time out of a $d$-dimensional one, which could be any, even infinite? 

i) In this paper we answer the question of how the {\em internal space} may result 
to restrictions on the choice of the signature of space-time   in 
any $d$-dimensional space, assuming  {\em the
Hermiticity of the equations of motion operator, its linearity in the $d$-momentum $p^a$ and the 
irreducibility of the Lorentz group representation in the} {\em internal space}.
 All these assumptions seem very mild:
In standard quantum mechanics, Hermiticity of the Hamiltonian and thereby of the equations of motion operator 
guarantees a real value for the energy, 
unitarity in the time-development of a system, and conservation of probability.
It has been proven\cite{bojannorma2001} that in even-dimensional spaces of any $d$ all 
massless particles,
 respecting the Poincar\' e symmetry, obey equations of motion, which are linear in the $p^a$-momentum.
The authors of ref.\cite{siegelzwiebach} inform us that, starting from  conformal 
symmetry, they came to the same conclusion for
any dimension. 
The requirement that solutions of the equations of motion operator should be a linear 
superposition of the minimum number of basic states of the  Lorentz group 
 possible leads to a choice of operators which the operator of equations of motion should 
commute with.  We select the operator of handedness,  which is the Casimir of
the Lorentz group in any dimensional space. Massless particles, obeying the 
equations of motion, are  in any dimensional space either left- or right-handed. 
(Since coefficients are complex numbers, 
we shall call these representations complex representations.)
Handedness seems to  play a fundamental role in Nature: one of the assumptions
of the Standard model, dictated by experiment, is that only massless spinors of left handedness
carry a weak charge, while right handed fermions are weak chargeless. Because of this 
property, spinors remain almost massless on the Planck scale. In other words, the Dirac equation
for massive  particles can only have solutions within the space of left- and right-handed states.  
According to the Standard model, spinors receive small (observable) masses by interacting with 
the Higgs fields, which (by 
assumption) give them masses of the order of only the weak scale. This is known as the mass 
protection mechanism.

ii) In this paper we draw attention to a fascinating  property that the mass protection 
mechanism only occurs in even-dimensional spaces.  In  odd-dimensional spaces, solutions of
equations of motion span the same space for massless particles as they do for massive particles. Therefore,
the procedure of excluding part of the Hilbert space does not work, and 
accordingly no mass protection mechanism can occur. All spinors could accordingly 
acquire large masses - say the Planck mass. No interaction with for instance the Higgs is required to 
assure them a mass. But we also have no reason  to believe that this mass is a verry small one in 
comparison with the Planck scale.  
Spinors in odd dimensional spaces could be invisible at ''low'' - experimentally accessible - energies.

iii) In this paper we argue that the stability of the equations of motion 
against a possible break of the Lorentz invariance privileges four-dimensional space (in addition to  
the two-dimensional one). (We are not saying that other dimensions are excluded.) 

We  consider only free (non interacting) fields. Apart from  momentum 
degrees of freedom, we  also consider spin  degrees of freedom and, for the sake of simplicity and 
transparency of  presentation of our proofs and of the consequences of the proofs, we  
treat only spinors. For a general case of any spin, the reader should see\cite{holgernorma2000}. 
We would however like to point out that, using the
Bargmann-Wigner prescription\cite{bargmannwigner},  any spin state can be constructed 
out of spinor states and accordingly all the requirements regarding the signature of the metric presented
for spinors should be in agreement with the requirements of all other spin particles.	

The ''working hypotheses'' of the paper is that the fundamental space is $d$-dimensional, with $d$ being 
any integer number, perhaps even infinite. In this paper we seek arguments which might have ''led'' Nature 
to ''end up'' with (''mostly'', that is effectively,)
four-dimensional space-time, with one time and three space dimensions. We do not discuss  any mechanism which
would lead from some large (or even infinite) dimension down to four dimensions. Such a mechanism is 
certainly needed in theories like 
string theories and Kaluza-Klein theories, as well as in the approach taken by one of us, which assumes that spins 
in $d$-dimensional space manifest in four-dimensional subspace as spin and all the known charges. We also 
do not claim that the proposed assumptions and heir argued consequences are  the only reasons for
the ''chosen'' dimension and signature of our observed world. But since the assumptions are rather mild ones, we 
expect the conclusions of this paper to be valid for any theory. Although we treat only massless 
noninteracting fields (of any spin), we expect that interactions among fields will not alter the conclusions,
at least not for those types of interaction which can be switched on perturbatively. Further assumptions could,
of course, limit further the allowed signatures.

\section{Equations of motion} \label{equationsofmotion}

 Massless spinors, if they preserve the Poincar\' e symmetry, obey  {\em equations of motion, which are 
 linear in the momentum $d$-vector} as proven in refs.\cite{bojannorma2001,siegelzwiebach}. 
We shall use  somewhat generalized equations of motion of the type 
\begin{eqnarray}
(f^a(S^{cd}) \;p_a) |\psi> 
= 0,
\label{lineqf}
\end{eqnarray}
with $a=0,1,2,3,5,\cdots,d$ 
where $\;f^a$ is for each $a$ any function of the generators of the Lorentz transformations $S^{ab}$
in the {\em internal space} of spin degrees of freedom. 

The total generators of the Lorentz 
transformations would read $M^{ab} = L^{ab} + 
S^{ab}$, with $L^{ab} = x^a p^b - x^b p^a$, which is the generator of the Lorentz transformations in
ordinary space. In order for the equations of motion operator to be linear in $p^a$, $f^a$ can depend only
on $S^{ab}$, that is on ''internal space''. Both $L^{ab}$ 
and  $S^{ab}$, as well as their sum,  fulfil the Lorentz algebra: 
$\;[M^{ab}, M^{cd}]_-  = -i(\eta^{ac} M^{bd} + \eta^{bd} M^{ac} - \eta^{ad} M^{bc} - \eta^{bc} M^{ad}),$ where
$\eta^{ab}= diag\{\eta^{00},\eta^{11},\cdots,\eta^{dd} \}$ is a (not yet specified) metric tensor
with $(\eta^{aa})^2 =1, $ for each $a$. (In four-dimensional space the two well known linear equations of motion
in the momentum 
for free fields are the Dirac (or equivalently the Weyl) equation for spin $\frac{1}{2}$ fermions (spinors) and the 
Maxwell equations for gauge Yang-Mills of spin $1$ fields.)

Since this paper considers only spinors, the generators $S^{ab}$, which  for spinors also fulfil  
the equation $\{S^{ab}, S^{ac}\}_+ = \frac{1}{2}\eta^{aa}\eta^{bc}$, can be expressed in terms of the
operators $\gamma^a, \; a=0,1,2,3,5,\cdots,d,$ (operating again in ''internal space'')  fulfilling 
the Clifford algebra
\begin{eqnarray}
\{\gamma^a,\gamma^b \}_+ := \gamma^a \gamma^b + \gamma^b\gamma^a = 2 \eta^{ab}
\label{cliff}
\end{eqnarray}
as follows
\begin{eqnarray}
S^{ab} = \frac{i}{4} [\gamma^{a},\gamma^b]_- := \frac{i}{2} \{\gamma^a \gamma^b -\eta^{ab}\}.
\label{gammasab}
\end{eqnarray}
The generalized linear equations can then be written in the form
\begin{eqnarray}
{\cal D}(\gamma^b) \;\gamma^a p_a =0.
\label{lineq}
\end{eqnarray}
(The reader should recall that the Dirac equation for massless spinors is usually written in the form 
$\; \gamma^a p_a =0.$)
{\em The Hermiticity condition}  reads
\begin{eqnarray}
\gamma^{a\dagger}({\cal D}(\gamma^b))^{\dagger} = {\cal D}(\gamma^b) \gamma^a,
\label{hereq}
\end{eqnarray}
since the operator $p^a$ is a Hermitean one, if the usual inner product in ordinary space is assumed.
The requirement of Hermiticity without allowing for an extra {\em internal space} matrix ${\cal D}(\gamma^b)$
would be too strong a requirement.
Performing Hermitean conjugation of Eq.(\ref{cliff}) and requiring   that the inner product
of a ket $\gamma^a|\psi>$ and a bra $(\gamma^a|\psi>)^{\dagger}$ has to have the same value as $<\psi|\psi>$ 
which leads to the unitarity condition   
$\gamma^{a\dagger} \gamma^a = I$, for any $a$,  we find
\begin{eqnarray}
\gamma^{a\dagger}= \eta^{aa}\gamma^a 
\label{gammaher}
\end{eqnarray}
and accordingly $ S^{ab\dagger} =  \eta^{aa}\eta^{bb} S^{ab}$.
According to Eqs.(\ref{gammaher},\ref{hereq}) the Hermiticity condition for the equations of 
motion operator reads
\begin{eqnarray}
({\cal D}(\gamma^b))^{\dagger} = \gamma^a {\cal D}(\gamma^b) \gamma^a,\quad {\rm for\;each\;a}.
\label{hereqd}
\end{eqnarray}
We define, according to refs.\cite{norma93,bojannorma2001}, the operator $\Gamma$, which in even-dimensional
spaces determines  the handedness of states for any spin. In this paper we shall express the operator of
handedness in terms of $\gamma^a$'s, since we treat only spinors ($\Gamma$ is for $d=4$ and for spinors
known as $\gamma^5$). It has then meaning for any dimensional 
space 
\begin{eqnarray}
\Gamma &=& \prod_{a} \; (\sqrt{\eta^{aa}}\;\gamma^a) \cdot \left\{ \begin{array}{rr}
(i)^{\frac{d}{2}}, \quad \hbox{for} \quad d \quad \hbox{even}\\
(i)^{\frac{d-1}{2}}, \quad \hbox{for} \quad d \quad \hbox{odd}
\end{array} \right.
\label{ggamma}
\end{eqnarray}
and the product of $\gamma^a$'s is assumed to be in the rising order with respect to index $a$.
We chose the phase such that the operator $\Gamma$ is Hermitean and its square is the unit operator
\begin{eqnarray}
\Gamma ^+&=& \Gamma, \quad \Gamma^2 = I.
\label{ggammaher}
\end{eqnarray}
We then easily find that
\begin{eqnarray}
\{\Gamma, \gamma^a \}_{\pm} &=& \left\{ \begin{array}{rr}
0, \;\; \hbox{for} \; + \; \hbox{sign} \;\;\hbox{and} \;\;  d \;\; \hbox{even}\\
0, \;\; \hbox{for} \; - \; \hbox{sign} \;\;\hbox{and} \;\;  d \;\; \hbox{odd}
\end{array} \right.
\label{ggammac}
\end{eqnarray}
and that $\Gamma$ is a Casimir of the Lorentz group, i.e. $\{ \Gamma, S^{ab}\}_- = 0.$

\section{Reducibility of representations} \label{reducibilityofrepresentations}

Eqs.(\ref{lineq}) and (\ref{hereqd}) concern the linearity and Hermiticity requirements for the operator of 
equations of motion. Only {\em the reducibility} has, according
to our assumptions, yet to be taken into account. We accordingly require, that the operator of equations of
motion and the operator of handedness commute
\begin{eqnarray}
\{\Gamma, {\cal D} \gamma^a p_a \}_- \;=\; \{\Gamma, {\cal D} \; \gamma^a \}_-\;=\; 0.
\label{ggammared}
\end{eqnarray}
The last equation has to be fulfilled for each $a$. We multiply  this  equation  
from the left by $\Gamma$ and take into account of the properties of $\Gamma$ ($\Gamma^+ = \Gamma$ and $\Gamma^2 = I$, 
Eq.(\ref{ggammaher})). It follows then, if  we also take into account Eqs.(\ref{ggammac}) (which states that in 
even-dimensional spaces $\Gamma$ anticommutes with $\gamma^a$'s, while in odd-dimensional spaces they commute) that
\begin{eqnarray}
{\cal D} \gamma^a + \Gamma^{-1} {\cal D} \Gamma \gamma^a  \;&=&\;0,\;\;{\rm for}\; 
d \;{\rm even},
\nonumber\\
{\cal D} \gamma^a -  {\cal D} \gamma^a  \;&=&\;0,\;\;{\rm for}\; 
d \;{\rm odd}.
\label{ggammared1}
\end{eqnarray}

{\em We conclude that in odd-dimensional spaces the reducibility requirement leads to no limitation whatsoever on
the signature of the metric.}
 We find that 
$ \Gamma^{-1} {\cal D}\Gamma$ = $(\gamma^d)^{-1}(\gamma^{d-1})^{-1}\cdots (\gamma^0)^{-1}{\cal D}\gamma^0\cdots
\gamma^{d-1}\gamma^d$. Using (\ref{hereqd}) and its Hermitean conjugate (${\cal D} = \gamma^a {\cal D}^{\dagger}
\gamma^a$)
we find by repetition that $\; \Gamma^{-1} {\cal D}\; \Gamma ={\cal D}\; \prod_{a}\; \eta^{aa}$. In order to fulfil 
Eq.(\ref{ggammared1}) for even
$d$ it follows that $\prod_{a}\; \eta^{aa} = -1.$
 
{\em In even-dimensional spaces the requirement is severe:  Solutions of equations of 
motion can only have well defined handedness in spaces of odd-time and odd-space signatures}. In a four-dimensional 
space it means that only {\em one time and three space signature is possible} (or the reverse, although this is 
not important, as the overall sign is not important). 

In four dimensions our result, $q$ being odd, means that we must have either $3$ time dimensions and one space dimension 
or, as we know, we have the reverse.

\section{Massless and massive solutions and the mass protection mechanism}\label{masslessandmassive}

We shall prove that the so called mass protection mechanism occurs only in even-dimensional spaces.
To prove this, we  look at the properties w.r.t. irreducibility of 
an equation  of motion operator with  a mass term as follows
\begin{eqnarray}
(\gamma^a p_a -m)|\psi> =0.
\label{masseq}
\end{eqnarray}
This is for $d$ equal to four and for the Minkowski metric the well-known Dirac equation, with an operator  
which is not Hermitean. If we want the operator to be Hermitean, we
multiply it by an appropriate matrix $D$ as above (which in the Dirac equation case is $\gamma^0$). 
We assume the equation to be valid for any dimension $d$.
Of course, for the mass-protection discussion we cannot retain the assumption of linearity, since the
mass term is obviously not linear in momentum. However, we do insist on the irreducibility assumption.
Let us check whether the irreducibility requirement is at all possible with the mass term added.
We use the fact that irreducible representations are obtained for the Dirac equation by projection 
with matrices $\frac{1}{2}(1\pm \Gamma)$ where $\Gamma$ is given by (\ref{ggamma}).
I.e. we restrict the state space of the Dirac equation for any $d$ and any signature to those states that obey
\begin{equation}
D(\gamma^ap_a -m)\frac{1}{2}(1-\Gamma)|\psi> =0,
\label{pd}
\end{equation}
$\Gamma |\psi> = |\psi>$
say, so that $|\psi> = \frac{1}{2}(1-\Gamma )|\psi>$.

For the projected Dirac equation to be irreducible, we must require that the equation (\ref{pd}) 
maps into the same subspace to which $\frac{1}{2}(1-\Gamma)|\psi>$  belongs. 
The requirement of these operators mapping onto the space projected on 
by $\frac{1}{2}(1-\Gamma)$ can be expressed by the requirement that projection by
the projection operator $\frac{1}{2}(1+\Gamma)$ on the orthogonal space will give zero
regardless of the state $|\psi>$. We accordingly require
\begin{equation}
\frac{1}{2}(1+\Gamma)D(\gamma^a p_a - m)\frac{1}{2}(1-\Gamma) =0
\label{noteq}
\end{equation}
for all values of $p_a$, on the operator level (which means that Eq.(\ref{noteq}) is not the 
equation of motion here) which  leads to
\begin{equation}
[\Gamma,Dm]=0 \quad {\rm and} \quad	 [\Gamma, D\gamma^a] =0.
\end{equation}
Combining these two equations with Eqs.(\ref{ggammac}) we see that it is impossible to be satisfied 
in even dimensions $d$.
In even dimensions the requirement of irreducibility prevents the mass term to occur 
since the only way out is to take the mass $m=0$. In the case of odd dimensions,  any mass is allowed
even after $\Gamma$-projection. This prevention of mass - in even $d$ -  is what is called 
mass protection, in the sense that a theory if is arranged so that its symmetries ( charges etc.)
enforce only the $\Gamma$-projected state space to be used then that theory can explain 
why the particles in question are massless, since a theory does not allow solutions of the 
equations of motion for massive spinors to exist, because the part of the space which has
opposite handedness is missing. In the Standard model\cite{standardmodel} 
the Weinberg-Salam-Higgs 
field at the end gives (by ``hand'') most fermions a ``little'' mass by breaking the gauge symmetries 
which caused the mass protection.

In odd dimensions $d$, however, it is not possible to  prevent spinors from acquiring a mass, 
because  
non-zero mass is allowed even when the $\Gamma$-projection is performed: the massive and massless 
Dirac equation have the solution within the same space of states. (Only the coefficients change but none
of the states disappear.) Taking the  
point of view that all parameters, say the mass, not forbidden are present with 
a scale of size given by an order of magnitude of a fundamental scale assumed to be very 
big - say the Planck scale - compared to energies per particle accessible in practice,
we conclude that in odd dimensions all spin one half particles will for practical
purposes acquire such large masses that they effectively can not be observed. {\em  So odd dimensions 
deviate from even ones by typically having all the masses of the fundamental scale,
while in even dimensions  the mass protection mechanism is possible }.

\section{Stability of the equations of motion leading to dimensions being at most 4}\label{stabilityofthe}

We shall add to our assumptions about the equations of motion operator in 
sections (\ref{reducibilityofrepresentations}) and (\ref{equationsofmotion})
the requirement that the equations of motion should be stable in the sense that if we {\em infinitesimally
destroy the Lorentz invariance} by adding a small extra term  still obeying the other 
assumptions, this term would in reality not disturb the equations of motion in the 
following sense: We could transform it away by shifting the coordinatization of the 
momentum $p_a$ and changing the metric tensor $\eta^{ab}$ into a new set of values.
This argumentation is really a rewriting of the old argument of ''Random Dynamics ''
of one of us \cite{holger}.  

Let us in fact imagine that the equations of motion (\ref{lineq}) with abstract 
notation $f^a = {\cal D} \gamma^a$ are modified  slightly  from
$f^ap_a|\psi>=0$
into 
\begin{equation}
(f^a  + f'^a)p_a |\psi>=0.
\label{modified}
\end{equation}
Can we then pretend that this new equation is indeed just of the same form as before,
but with slightly changed notation for the way one expands the momentum on  the basis vectors 
and the metric tensor $g^{ab}$ known from general relativity? In other words, can we 
by changing the basis for the $d$-momentum $p_a$, write the modified equations in a  
slightly more general form
\begin{equation}
g^{ab}f_ap_b |\psi>=0\;\;\; ?
\end{equation}
By counting degrees of freedom, we see that this could only be the case for a 
general modification term $\; f'^ap_a$ provided we have at most 4 dimensions, i.e. it is
at least needed that $d\le 4$.

This counting goes as follows: the number of degrees of freedom of the extra term $f'^ap_a $
is that of $d$ matrices with the number of columns and rows equal to the dimension
of the irreducible representation of the ''Weyl'', which is $2^{d/2 -1}$ for an even 
dimension $d$ and $2^{(d-1)/2}$ for an odd dimension $d$. Altogether, this means that the 
number of real parameters in the modification is $\; d \cdot 2^{d-2}$ for even  and
$\; d \cdot 2^{d-1}$ for  odd $d$. These modification parameters should be compensated by
$d(d+1)/2$ parameters in the metric tensor $g^{ab}$ (or $\eta^{ab}$) and $d(d-1)/2$ parameters 
associated with making a Lorentz transformation of the basis for the $d$-momentum
$p_a$. We need accordingly the inequality
\begin{equation}
d(d+1)/2 + d(d-1)/2 \ge d\cdot  2^{d-2}
\end{equation}
for even $d$, and
\begin{equation}
d(d+1)/2 + d(d-1)/2 \ge d\cdot 2^{d-1}
\end{equation}
for odd $d$.
These equations reduce to $d \ge 2^{d-2}$ and $ d \ge 2^{d-1} $
for even and odd $d$ respectively. Thus we must for even $d$ have $d \le 4$, while for
odd $d$ only $d=1$ satisfies the inequality. 

Accordingly, this stability requirement can only work for $d = 1,2$ and $4$ dimensions.

\section{Conclusion} \label{conclusion}

In this paper we sought arguments which might have ''guided'' Nature to ''mostly'' (effectively) manifest
in four-dimensional ordinary space, with one time and three space coordinates, in addition to 
the ''internal space''
of spins and charges. We argued that it is the
 ''internal space'' which forced Nature to ''make a choice'' of $d$ equal four, if  
rather mild assumptions about the properties of the operator of equations of motion for spinors are the 
meaningful ones we believe they are. (Further assumptions would further limit possible signatures
in $d$-dimensional space. While a somewhat relaxed stability assumption could tell, for example, why 
spin degrees of freedom in higher than four dimensions might demonstrate as (conserved) charges
in four-dimensional space-time.) These mild assumptions lead to 
various predictions concerning the space-time dimension\underline{s}. The five assumptions which 
we used were:

\begin{enumerate}

\item[1)] Linearity in momentum,

\item[2)] Hermiticity,

\item[3)] Irreducibility,

\item[4)] Mass protection,

\item[5)] Stability.

\end{enumerate}

We proved that, if we apply assumptions 1), 2) and 3)
in {\em even-dimensional spaces only odd-time and odd-space dimensions  are possible}, while in {\em odd-dimensional 
spaces all signatures are possible}. However, while in {\em even-dimensional spaces 
limitation to only one
irreducible representation, for example  left handedness,
enables fermions to remain  massless},
{\em this is not true for odd-dimensional spaces}, since the solutions for a massless and a massive case span the
same space. So applying assumption 4) we {\em exclude an odd number of space plus time dimensions}.

Applying assumption 1), 2), 3) and 5), we concluded that the total dimension should be $1,2,$ or $4$.
Taking account of assumption 4), we exclude $d=1$ and using the odd time and odd space 
dimensions we finally obtain from 
all our five assumptions only the  time-space dimensions $1+1,$ or $1 + 3$  ( or opposite). This is thus 
close to explaining the experimental numbers $1+3$.

We understand the time-space dimension $1+3 $ as an effective dimension.

Further studies along these lines might involve consideration of different kinds of representations, like   
Majoranas (the paper by the two authors of this paper, entitled ''The internal space is making 
the choice of the signature of space-time'', which also considers Majoranas,
is 
almost ready for publication), 
which are representations with real coefficients (i.e. using the field of real numbers). It turns, however, out  
that these kinds of representations have no mass protection
mechanism either, and the corresponding fields are accordingly invisible at low energies.

This paper treats only free (non interacting) fields of any spin in $d$-dimensional space of any signature
and any $d$. However, we expect that interactions among fields will not alter the conclusions of these paper, 
that is they do not result in additional limitations on possible signatures, if only those types of interactions
which can be switched on perturbatively are assumed.

Allowing any dimension $d$ with ordinary and spin degrees of freedom, we only look for properties of 
equations of motion, which could possibly be responsible for four 
ordinary effective (detectable) dimensions - in addition to ''internal space''. We did not consider 
mechanisms, which could have brought $d$ ordinary (and internal) dimensions to four effective 
dimensions, or 
the possibility (proposed by the approach of one of the authors\cite{norma92,norma93,norma01} of this paper) that  
spins of higher than four dimensions are responsible for charges in the effective $(1+3)$ dimension. 
We also do not comment on where the
limitation that only  states of one handedness (which leads to mass protection mechanism for even-dimensional
spaces) could come from.

(We have here by the number of times meant the number of dimensions with a certain signature of the metric. 
But there is another way in which one could define something that with some right could be called the
number of time dimensions. One could namely consider more than one equation of motion per field component. 
By in the present article considering only one equation of motion per field component, we have in this different
sense assumed just one time dimension. From that point of view one could even say that we put in a different 
meaning that here we have just one and thus especially an odd number of times.)

Yet the results of our conclusions, if the (rather mild) assumptions can be taken seriously (which we believe they
should), are conclusive. They are also restrictive for theories with additional degrees of freedom, such as  
string theories and Kaluza-Klein theories.

\section{Acknowledgement } 

This work was supported by Ministry of Education, 
Science and Sport of Slovenia and Ministry of Science of Denmark.  The authors would like to thank
to the communicator and his referees for stimulating suggestions to further clarify the paper.

\end{document}